\documentclass{ws-mpla}

\newcommand{\tbelrob}{\overset{\scriptscriptstyle BR}{\mathcal{T}}}
\newcommand{\tbelrobtype}{\overset{\scriptscriptstyle BR}{T}}
\newcommand{\tbel}{\overset{\scriptscriptstyle B}{\mathcal{T}}}
\newcommand{\tbeltype}{\overset{\scriptscriptstyle B}{T}}

\newcommand{\be}{\begin{equation}}
\newcommand{\ee}{\end{equation}}
\newcommand{\bea}{\begin{eqnarray}}
\newcommand{\eea}{\end{eqnarray}}

\newcommand{\fsymbtwo}{f}
\newcommand{\fsymbthree}{h}

\newcommand{\killingtwo}{K}
\newcommand{\killingthree}{L}

\newcommand{\confkillingtwo}{P}

\newcommand{\devcov}{\nabla}

\newcommand{\jbelrobfk}[1]{j^{#1}(\tbelrobtype,\:\fsymbtwo,\:\xi)}

\newcommand{\jbelrobfck}[1]{j^{#1}(\tbelrobtype,\:\fsymbtwo,\:\zeta)}

\newcommand{\jbelrobf}[1]{j^{#1}(\tbelrobtype,\:\fsymbthree)}
\newcommand{\Jbelrobf}[1]{J^{#1}(\tbelrobtype,\:\fsymbthree)}

\newcommand{\jbelrobk}[1]{j^{#1}(\tbelrobtype,\:\killingthree)}

\newcommand{\jbelf}[1]{j^{#1}(\tbeltype,\:\fsymbthree)}

\newcommand{\jbelrobky}[1]{j^{#1}(\tbelrobtype,\:Y)}

\begin{document}

\markboth{Ovidiu Tintareanu-Mircea}
{f-symbols, Killing tensors and conserved Bel-type currents}

%%%%%%%%%%%%%%%%%%%%% Publisher's Area please ignore %%%%%%%%%%%%%%
\catchline{}{}{}{}{}
%%%%%%%%%%%%%%%%%%%%%%%%%%%%%%%%%%%%%%%%%%%%%%%%%%%%%%%%%%%%%%%%%%%

\title{F-SYMBOLS, KILLING TENSORS AND CONSERVED BEL-TYPE CURRENTS}

\author{\footnotesize OVIDIU TINTAREANU-MIRCEA}

\address{Institute for Space Sciences\\
Magurele, P.O. Box MG-23, Ro-077125 Bucharest, Romania\\
ovidiu@spacescience.ro}

\maketitle

\pub{Received (Day Month Year)}{Revised (Day Month Year)}

\begin{abstract}
In the framework of the General Relativity we show that from
three generalizations of Killing vector fields, namely f-symbols, symmetric
St\"{a}ckel-Killing and antisymmetric Killing-Yano tensors,
some conserved currents can be obtained through adequate
contractions of the above mentioned objects with
rank four tensors having the properties of Bel or Bel-Robinson tensors
in Einstein spaces.

\end{abstract}

\ccode{PACS Nos.: 04.20-q}

\section{Introduction}	
%
%
%
% Bel and Bel-Robinson
%
%
The Bel tensor
(the superenergy tensor of the gravitational field
[\refcite{Bonilla-Senovilla-1997}, \refcite{Senovilla-2000}])
was first introduced in
[\refcite{Bel-1959}]
by exploring an analogy between the gravitational and the electromagnetic fields.
A rank four tensor was obtained, an analogous of the energy-momentum tensor
for the electromagnetic field but with the Riemann tensor of curvature instead of the
electromagnetic field tensor
\bea
\tbel{}^{\mu\nu\lambda\rho}&\equiv&\frac{1}{2}
\left(
R^{\mu\alpha\lambda\beta}R^{\nu~\rho}_{~\alpha~\beta}+
\star R\star{}^{\mu\alpha\lambda\beta}+
\star R\star{}^{\nu~\rho}_{~\alpha~\beta}
\right.
\nonumber\\
&&
\left.
\star R^{\mu\alpha\lambda\beta}\star R^{\nu~\rho}_{~\alpha~\beta}+
R\star{}^{\mu\alpha\lambda\beta}R\star{}^{\nu~\rho}_{~\alpha~\beta}
\right),
\eea
where $R_{\mu\nu\lambda\rho}$ is the Riemann curvature tensor and the various
types of Hodge duals are
$\star R_{\mu\nu\lambda\rho}
\equiv
\frac{1}{2}\varepsilon_{\mu\nu\alpha\beta}R^{\alpha\beta}_{~~\lambda\rho}$,
$R\star_{\mu\nu\lambda\rho}
\equiv
\frac{1}{2}\varepsilon_{\lambda\rho\alpha\beta}R^{~~\alpha\beta}_{\mu\nu}$,
$\star R\star_{\mu\nu\lambda\rho}
\equiv
\frac{1}{4}\varepsilon_{\mu\nu\alpha\beta}\varepsilon_{\lambda\rho\nu\tau}R^{\alpha\beta\nu\tau}$,
$\varepsilon_{\mu\nu\lambda\rho}$
being the canonical volume element of the spacetime.
The Bel tensor is traceless $\tbel{}^{\alpha~\lambda\rho}_{~\alpha}=0$
and have the following symmetries
\be
\tbel{}^{\mu\nu\lambda\rho}=
\tbel{}^{(\mu\nu)(\lambda\rho)}=\tbel{}^{\lambda\rho\mu\nu}.
\ee
It's divergence is
$
\devcov_{\mu}\tbel{}^{\mu\nu\lambda\rho}
=
R^{\nu~\lambda}_{~\alpha~\beta}J^{\rho\alpha\beta}+
R^{\nu~\rho}_{~\alpha~\beta}J^{\lambda\beta\alpha}-
\frac{1}{2}g^{\lambda\rho}R^{\nu}_{~\alpha\beta\gamma}J^{\beta\gamma\alpha}
$
where
$
J_{\mu\nu\lambda}
\equiv
\devcov_{\mu}R_{\nu\lambda}-\devcov_{\nu}R_{\mu\lambda}
$,
$R_{\mu\nu}$ being the Ricci tensor.
It is therefore obvious that the Bel tensor is locally conserved in empty or Einstein spacetimes.

In vacuum the Bel tensor is known as Bel-Robinson tensor and it was introduced for the first time in
[\refcite{Bel-1958}].
The Bel-Robinson tensor is defined as
\be\label{bel-rob-def-01}
\tbelrob{}^{\mu\nu\lambda\rho}
=
R^{\mu\alpha\lambda\beta}R^{\nu~\rho}_{~\alpha~\beta}+
\star R^{\mu\alpha\lambda\beta}\star R^{\nu~\rho}_{~\alpha~\beta}
\mbox{ (in vacuum)}
\ee
and satisfies
\be\label{br-prop-vacuum-01}
\tbelrob{}^{\mu\nu\lambda\rho}=\tbelrob{}^{(\mu\nu\lambda\rho)},
\quad
\tbelrob{}^{\mu~\lambda\rho}_{~\mu}=0.
\ee
In vacuum and in Einstein spaces the Bel-Robinson is divergenceless
\be\label{br-prop-div-vac}
\devcov_{\mu}\tbelrob{}^{\mu}_{~\nu\lambda\rho}=0.
\ee
A tensor with the properties (\ref{br-prop-vacuum-01}) and (\ref{br-prop-div-vac})
of the Bel-Robinson tensor can be constructed for any spacetime, irrespective of it's emptiness character
if in (\ref{bel-rob-def-01}) the conformal Weyl tensor is used instead of the Riemann curvature tensor
\be
\tbelrob{}^{\mu\nu\lambda\rho}
\equiv
C^{\mu\alpha\lambda\beta}C^{\nu~\rho}_{~\alpha~\beta}+
\star C^{\mu\alpha\lambda\beta}
\star C^{\nu~\rho}_{~\alpha~\beta}.
\ee
A study of the Bel-Robinson tensor and of its impact on the evolution of the Universe
within the framework of the Bianchi type-I spacetime can be found in
[\refcite{Saha-Rikhvitsky-Visinescu-MPLA-21-2006}].

In the noticeable cases when Bel and Bel-Robinson tensors and generally the superenergy tensors
of physical fields are divergenceless various conserved currents can be built
in the presence of Killing vector fields
[\refcite{Senovilla-2000},
\refcite{Lazkoz-Senovilla-Vera-ERE}].

When we have interacting fields such as in Einstein-Klein-Gordon or Einstein-Maxwell theories,
the superenergy currents associated with individual fields are not conserved
and an interchange of some superenergy quantities take place between the gravitational and
the non-gravitational fields in such a manner that a total, mixed current is conserved.
For example, it was shown
[\refcite{Senovilla-2000}, \refcite{Senovilla-MPLA-2000}]
that for Einstein-Klein-Gordon theory a mixed conserved superenergy
current can be constructed whenever there is also a Killing vector field
\be\label{cons-curr-comb}
\devcov^{\mu}((\tbel{}_{\mu\nu\lambda\rho}+S_{\mu\nu\lambda\rho})\xi^{\nu}\xi^{\lambda}\xi^{\rho})=0,
\ee
where $S_{\mu\nu\lambda\rho}$ is the superenergy tensor of the scalar field.
Moreover, this current reduces to the corresponding Bel conserved current in the absence
of the scalar field and to the conserved superenergy current of the scalar field in a flat spacetime if the
gravitational field is removed. Such conservation laws makes the superenergy tensors not only
mathematically appealing, they opening perspectives in identifying their physical significance.

There are also situations when the conserved currents for individual fields
exists under very general circumstances even if the divergenceless property does not hold.
In
[\refcite{Lazkoz-Senovilla-Vera-CQG-2003}]
it was shown that independently conserved
currents can be constructed from the Bel tensor when there is a hypersurface orthogonal
Killing vector or when there are two commuting Killing vectors that acts orthogonally
transitive on non-null surfaces.
In the first case we have a current proportional with the Killing vector
\be\label{cons-curr-prop}
\tbel{}_{\mu\nu\lambda\rho}\xi^{\nu}\xi^{\lambda}\xi^{\rho}=\gamma\xi_{\mu},
\quad
\devcov^{\mu}(\tbel{}_{\mu\nu\lambda\rho}\xi^{\nu}\xi^{\lambda}\xi^{\rho})=0,
\ee
while in the second we have four currents lying in the 2-plane generated by the two Killing vectors
\be\label{cons-curr-2-plane}
\tbel{}_{\mu(\nu\lambda\rho)}\xi_{i}^{\nu}\xi_{j}^{\lambda}\xi_{k}^{\rho}
=
\alpha_{ijk}\xi_{1\mu}+\beta_{ijk}\xi_{2\mu},
\quad
\devcov^{\mu}(\tbel{}_{\mu(\nu\lambda\rho)}\xi_{i}^{\nu}\xi_{j}^{\lambda}\xi_{k}^{\rho})=0.
\ee

The case of a four-dimensional Einstein-Maxwell theory with a source-free
electromagnetic field is discussed in
[\refcite{Eriksson-CQG-2006}, \refcite{Eriksson-CQG-2007}].
While in
[\refcite{Eriksson-CQG-2006}]
it is shown that for electromagnetic fields
inheriting the symmetry of the spacetime the Chevreton superenergy
currents are individually conserved if there are hypersurface orthogonal Killing vector fields, in
[\refcite{Eriksson-CQG-2007}]
is treated the case when there is a two-parameter Abelian isometry
group that acts orthogonally transitive on non-null surfaces and it is shown that the corresponding
superenergy currents lies in the orbits of the group and are conserved.

As can be seen from (\ref{cons-curr-comb}), (\ref{cons-curr-prop}) and (\ref{cons-curr-2-plane})
a Killing vector field is required for constructing Bel type currents.
There are many generalizations of Killing vector fields to higher rank tensors,
the most notable being St\"{a}ckel-Killing tensors, Killing-Yano tensors and f-symbols.
In the next section we show that when a spacetime admits one or more of the above geometrical objects
generalizing a Killing vector field, some conserved currents can be constructed from
Bel or Bel-Robinson type tensors.
%
%
%
%%%%%%%%%%%%%%%%%%%%%%%%%%%%%%
\section{Conserved currents} %
%%%%%%%%%%%%%%%%%%%%%%%%%%%%%%
%
%
%
We present three types of generalization of Killing vector fields -
f-symbols, Killing-Yano and St\"{a}ckel-Killing tensors -
and use them to construct Bel-type conserved currents.
%
%
%
%%%%%%%%%%%%%%%%%%%%%%%%%%%%%%%%%%%%%%%%%%%%%%%
\subsection{f-symbols and conserved currents} %
%%%%%%%%%%%%%%%%%%%%%%%%%%%%%%%%%%%%%%%%%%%%%%%
%
%
%
f-symbols have been introduced in
[\refcite{Gibbons-Rietdijk-vanHolten-NPB-404-1993}]
where they naturally arises in investigation of extra supersymmetries
in a pseudo-classical model for spinning particles.
A study about their existence for a particular spacetime can be found in
[\refcite{Cotaescu-Visinescu-f-symb}].
A rank two f-symbol is a tensor field $\fsymbtwo{}_{\mu\nu}$ with no particular
symmetries satisfying
\be\label{f-symb-r2-def}
\devcov_{\lambda}\:\fsymbtwo{}_{\mu\nu}+\devcov_{\nu}\:\fsymbtwo{}_{\mu\lambda}=0.
\ee
Eq. (\ref{f-symb-r2-def}) is identical with the equation for Killing-Yano tensors,
what makes the difference being the antisymmetry property not required for f-symbols.
It follows that the divergence in the second index vanishes
$
\devcov_{\nu}\:\fsymbtwo{}_{\mu}^{~\nu}=0
$
while for the divergence in the first index we have
$
\devcov_{\mu}\:\fsymbtwo{}_{~\nu}^{\mu}=-\partial_{\nu}\:\fsymbtwo{}_{\mu}^{~\mu}
$,
i.e. the first index divergence vanishes if the trace of the
f-symbol is a constant.
In this particular case, since the metric tensor is covariantly constant
and trivially satisfies equation (\ref{f-symb-r2-def}), we can substract
the trace part from the f-symbol and the remaining trace free part
is itself a f-symbol.

The symmetric part of a f-symbol
$S_{\mu\nu}=\frac{1}{2}(\fsymbtwo{}_{\mu\nu}+\fsymbtwo{}_{\nu\mu})$
is a St\"{a}ckel-Killing tensor, i.e. it satisfies
$
\devcov_{(\lambda}S_{\mu\nu)}=0
$,
while for the antisymmetric part $A_{\mu\nu}=\frac{1}{2}(\fsymbtwo{}_{\mu\nu}-\fsymbtwo{}_{\nu\mu})$ we have
$
\devcov_{\nu}A_{\mu\lambda}+\devcov_{\lambda}A_{\mu\nu}=\devcov_{\mu}S_{\nu\lambda}
$,
meaning that the antisymmetric part satisfies itself the f-symbol equation
(\ref{f-symb-r2-def}) (i.e. it is a Killing-Yano tensor) if the symmetric part
$S_{\mu\nu}$ vanishes or is covariantly constant.

By analogy with the integrability equation
\be\label{int-cond-k-v}
\devcov_{\nu}\devcov_{\lambda}\xi_{\mu}=-R^{\alpha}_{~\nu\lambda\mu}\xi_{\alpha}=0
\ee
satisfied by any Killing vector field
$\xi_{\mu}$,
a similar equation can be writen down for an arbitrary f-symbol.
For example, for a rank two f-symbol, if the defining equation
($\devcov_{\lambda}f_{\nu\mu}+\devcov_{\nu}f_{\lambda\mu}=0$)
is covariantly derived once again we obtain
\be\label{f-symb-eq-der-001}
\nabla_{\lambda}\nabla_{\rho}f_{\mu\nu}+\nabla_{\lambda}\nabla_{\mu}f_{\rho\nu}=0.
\ee
After we rewrite twice the above equation with
$\lambda\rightarrow\rho\rightarrow\mu\rightarrow\lambda$
and add up the resulting equations (the last one with a minus sign)
then make use of the Ricci
($[\nabla_{\lambda},\:\nabla_{\rho}]f_{\mu\nu}=
-R_{\lambda\rho\mu}^{~~~\alpha}f_{\alpha\nu}-R_{\lambda\rho\nu}^{~~~\alpha}f_{\mu\alpha}$)
and algebraic Bianchi
($R^{\mu}_{~\alpha\beta\gamma}+R^{\mu}_{~\gamma\alpha\beta}+R^{\mu}_{~\beta\gamma\alpha}=0$)
identities we end up with an integrability
condition for a f-symbol, analogous to the eq. (\ref{int-cond-k-v})
for the Killing vector fields
\be\label{int-cond-f-symb}
2\nabla_{\lambda}\nabla_{\rho}f_{\mu\nu}=
2R_{\lambda~\rho\mu}^{~\alpha}f_{\alpha\nu}+
R_{\nu~\rho\lambda}^{~\alpha}f_{\mu\alpha}+
R_{\nu~\lambda\mu}^{~\alpha}f_{\rho\alpha}+
R_{\nu~\rho\mu}^{~\alpha}f_{\lambda\alpha}.
\ee
Observe that the sum of the last three terms is not over
the circular permutation of $\mu,~\lambda$ and $\rho$ indices
and since $f_{\mu\nu}$ have no particular symmetries the above equation can not be further
compacted or simplified.

The first current we construct is from a rank-four tensor
$\tbelrobtype{}_{\mu\nu\lambda\rho}$
with the symmetry and divergence properties of the Bel-Robinson tensor in vacuum
or Einstein spaces
\bea
\tbelrobtype{}^{\mu\nu\lambda\rho}&=&\tbelrobtype{}^{(\mu\nu\lambda\rho)}\label{BR-type-symm}\\
\devcov_{\mu}\:\tbelrobtype{}^{\mu\nu\lambda\rho}&=&0\label{BR-type-div}\\
\tbelrobtype{}^{\alpha}_{~\alpha\mu\nu\lambda}&=&0\label{BR-type-trace}
\eea
and a rank three f-symbol
\be\label{f-symb-r3-def}
\devcov_{\rho}\:\fsymbthree{}_{\mu\nu\lambda}+
\devcov_{\lambda}\:\fsymbthree{}_{\mu\nu\rho}=0.
\ee
(The intrinsic algebraic characterization the Bel-Robinson type tensors was obtained in
[\refcite{Bergqvist-Lankinen-2004}],
where the necessary and sufficient
conditions for a rank four symmetric tensor to be a Bel-Robinson type tensor
- i.e. to be the superenergy tensor of a tensor with the same algebraic symmetries as the Weyl tensor -
are found.
This may be seen as the first Rainich theory result for rank four tensors.)

Contracting the above tensors we obtain the vector field
\be\label{j-3-def}
\jbelrobf{\mu}=\tbelrobtype{}^{\mu\nu\lambda\rho}\:\fsymbthree{}_{\nu\lambda\rho}
\ee
whose divergence is
\be\label{div-3-j-part}
\devcov_{\mu}\jbelrobf{\mu}=
\devcov_{\mu}\:\tbelrobtype{}^{\mu\nu\lambda\rho}\fsymbthree{}_{\nu\lambda\rho}+
\tbelrobtype{}^{\mu\nu\lambda\rho}\devcov_{\mu}\:\fsymbthree{}_{\nu\lambda\rho}=
\tbelrobtype{}^{\mu\nu\lambda\rho}\devcov_{\mu}\:\fsymbthree{}_{\nu\lambda\rho}.
\ee
Taking into account the defining equation (\ref{f-symb-r3-def}), the above divergence becomes
\be\label{sir-egal-001}
\devcov_{\mu}\jbelrobf{\mu}=
-\tbelrobtype{}^{\mu\nu\lambda\rho}\devcov_{\rho}\:\fsymbthree{}_{\nu\lambda\mu}=
-\tbelrobtype{}^{\rho\nu\lambda\mu}\devcov_{\rho}\:\fsymbthree{}_{\nu\lambda\mu}
\ee
the last equality being a consequence of the total symmetry property
$\tbelrobtype{}^{\mu\nu\lambda\rho}=\tbelrobtype{}^{(\mu\nu\lambda\rho)}$.
What it was obtained is $\devcov_{\mu}\jbelrobf{\mu}=-\devcov_{\mu}\jbelrobf{\mu}$
which means that the current $\jbelrobf{\mu}$ defined in (\ref{j-3-def}) is conserved.
The Gauss theorem for vector fields can be applied on a covariantly conserved current
since from a divergencefree vector field in a covariant sense
\be
\devcov_{\mu}\jbelrobf{\mu}=0
\ee
we can construct the following current
\be
\Jbelrobf{\mu}=(-g)^{\frac{1}{2}}\:\jbelrobf{\mu}
\ee
whose ordinary divergence vanishes
\be\label{null-ord-div-curr}
\partial_{\mu}\Jbelrobf{\mu}=0.
\ee
Eq. (\ref{null-ord-div-curr}) can be regarded as a conservation law for a fluid
whose density is $(-g)^{\frac{1}{2}}\jbelrobf{0}$ and whose motion is encapsulated in
$(-g)^{\frac{1}{2}}\jbelrobf{i}$, $i=1,2,3$.
The Gauss theorem can then be used to obtain an integral form and
if there is no flow through the boundary of the integration domain we have
\be
\int(-g)^{\frac{1}{2}}\jbelrobf{0}\:d^{3}x=ct.
\ee

Contracting the rank four tensor $\tbelrobtype{}^{\mu\nu\lambda\rho}$
with the symmetry, divergence and trace properties of the Bel-Robinson tensor from
eq. (\ref{BR-type-symm}), (\ref{BR-type-div}) and (\ref{BR-type-trace}) with a rank two f-symbol
$\fsymbtwo{}_{\mu\nu}$ we can obtain conserved currents, but only if the spacetime
admits symmetries generated by Killing or conformal Killing vector fields.
Thus we start with the rank four tensor
$\tbelrobtype{}^{\mu\nu\lambda\rho}$
with the above mentioned properties and a rank two f-symbol
$\fsymbtwo{}_{\mu\nu}$.
Because of the total symmetry properties of the Bel-Robinson tensor
there are only one independent contraction with the f-symbol $\fsymbtwo{}_{\mu\nu}$ namely
\be
T^{\mu\nu}=\tbelrobtype{}^{\mu\nu\alpha\beta}\:\fsymbtwo{}_{\alpha\beta}
\ee
which is evidently symmetric and easy to see that is also divergence- and trace-free
\be\label{br-f2-symm-div}
T^{\mu\nu}=T^{(\mu\nu)},\quad \devcov_{\mu}T^{\mu\nu}=0,\quad T^{\mu}_{~\mu}=0.
\ee
Even if the property $\devcov_{\mu}T^{\mu\nu}=0$ from (\ref{br-f2-symm-div})
represents a local conservation law, an integral form cannot be constructed
via Gauss theorem since the integrand would be a vector field and we cannot
add vectors at different points when we deal with curved spacetimes.
However, if the spacetime admit isometries (generated by Killing vector fields)
or conformal symmetries (generated by conformal Killing vector fields) we can go further
and construct a divergenceless vector field for which the Gauss theorem can be applied
to obtain an integral conservation law.

Thus if $\xi_{\mu}$ is a Killing vector field, than the current $\jbelrobfk{\mu}$ given by
\be\label{cons-curr-f-symb-2-br}
\jbelrobfk{\mu}=
T^{\mu\nu}\xi_{\nu}=
\tbelrobtype{}^{\mu\nu\alpha\beta}\:\fsymbtwo{}_{\alpha\beta}\xi_{\nu}
\ee
is covariantly conserved
\be
\devcov_{\mu}\jbelrobfk{\mu}=0.
\ee
Demonstration is simple, deriving from symmetry and divergence property of $T^{\mu\nu}$
and also taking into account the Killing vectors equation.

Since
$T^{\mu\nu}=\tbelrobtype{}^{\mu\nu\alpha\beta}\:\fsymbtwo{}_{\alpha\beta}$
is also traceless
(in four dimensions the Bel-Robinson tensor is traceless),
a conserved current can be constructed as
\be\label{j-br-f-conf}
\jbelrobfck{\mu}=
T^{\mu\nu}\zeta_{\nu}=
\tbelrobtype{}^{\mu\nu\alpha\beta}\:\fsymbtwo{}_{\alpha\beta}\zeta_{\nu}
\ee
where $\zeta_{\mu}$ satisfies
\be\label{conf-k-vect-def}
\devcov_{\nu}\zeta_{\mu}+\devcov_{\mu}\zeta_{\nu} = \chi g_{\mu\nu}
\ee
i.e. $\zeta_{\mu}$ is a conformal Killing vector field.
Since $\devcov_{\mu}T^{\mu\nu}=0$ and $T^{\mu\nu}=T^{\nu\mu}$
we have for the divergence of (\ref{j-br-f-conf})
\be
\devcov_{\mu}\jbelrobfck{\mu}=T^{\mu\nu}\devcov_{\mu}\zeta_{\nu}=T^{\mu\nu}\devcov_{\nu}\zeta_{\mu}.
\ee
Taken then into account the defining equation of conformal Killing vector fields (\ref{conf-k-vect-def})
we have
\be
\devcov_{\mu}\jbelrobfck{\mu}=T^{\mu\nu}\chi g_{\mu\nu}-T^{\mu\nu}\devcov_{\nu}\zeta_{\mu}
\ee
which because of the traceless property $T^{\mu\nu}g_{\mu\nu}=0$ lead to
\be
\devcov_{\mu}\jbelrobfck{\mu}=-T^{\mu\nu}\devcov_{\nu}\zeta_{\mu}=-\devcov_{\mu}\jbelrobfck{\mu}=0.
\ee

From a tensor
$\tbeltype{}^{\mu\nu\lambda\rho}$
with the symmetry and divergence properties of the Bel tensor in vacuum or Einstein spaces
\bea
\tbeltype{}^{\mu\nu\lambda\rho}&=&\tbeltype{}^{(\mu\nu)(\lambda\rho)}=\tbeltype{}^{\lambda\rho\mu\nu}\label{B-type-symm}\\
\devcov_{\mu}\:\tbeltype{}^{\mu\nu\lambda\rho}&=&0\label{B-type-div}\\
\tbeltype{}^{\alpha}_{~\alpha\mu\nu\lambda}&=&0\label{B-type-trace}
\eea
and a rank three f-symbol
$\fsymbthree{}_{\mu\nu\lambda}$
we can form a locally conserved current
$\jbelf{\mu}$.
Because of the symmetries of the Bel tensor there
are a unique independent contraction in three indices between
$\tbeltype{}^{\mu\nu\lambda\rho}$ and $\fsymbthree{}_{\mu\nu\lambda}$,
namely
\be\label{j-fsymb3-bel-def}
\jbelf{\mu}=\tbeltype{}^{\mu\nu\lambda\rho}\:\fsymbthree{}_{\nu\lambda\rho}
\ee
whose divergence is
\be
\devcov_{\mu}\:\jbelf{\mu}
=
\left(
\devcov_{\mu}\tbeltype{}^{\mu\nu\lambda\rho}
\right)
\fsymbthree{}_{\lambda\rho\nu}+\tbeltype{}^{\mu\nu\lambda\rho}\devcov_{\mu}\:\fsymbthree{}_{\lambda\rho\nu}
\ee
Because of
(\ref{B-type-div})
and using the defining equation
(\ref{f-symb-r3-def})
of a rank three f-symbol we have
\be
\devcov_{\mu}\jbelf{\mu}=
-\tbeltype{}^{\mu\nu\lambda\rho}\devcov_{\nu}\:\fsymbthree{}_{\lambda\rho\mu}=
-\tbeltype{}^{\nu\mu\lambda\rho}\devcov_{\nu}\:\fsymbthree{}_{\lambda\rho\mu}=
-\tbeltype{}^{\mu\nu\lambda\rho}\devcov_{\mu}\:\fsymbthree{}_{\lambda\rho\nu}
\ee
where we make use only of the symmetry
$\tbeltype{}^{\mu\nu\lambda\rho}=\tbeltype{}^{(\mu\nu)\lambda\rho}$ and then rename the '$\mu$'
and '$\nu$' indices. We have obtain that $\devcov_{\mu}\:\jbelf{\mu}=-\devcov_{\mu}\:\jbelf{\mu}$
which means the $\jbelf{\mu}$ current is divergenceless.
We don't make use of the symmetries of the Bel tensor related with the pairs of indices interchange.

Because of the symmetries of the Bel tensor there are only two independent
contractions between the Bel-type tensor and a rank two f-symbol $\fsymbtwo{}_{\mu\nu}$
\bea
T^{(1)\mu\nu}&=&\tbeltype{}^{\mu\nu\lambda\rho}\:\fsymbtwo{}_{\lambda\rho}\\
T^{(2)\mu\nu}&=&\tbeltype{}^{\mu\lambda\nu\rho}\:\fsymbtwo{}_{\lambda\rho}
\eea
The first contraction $T^{(1)\mu\nu}$ is obvious symmetric but has nonzero divergence
while the for the second contraction - which is not
symmetric - we have $\devcov_{\mu}T^{(2)\mu\nu}=0$ and $\devcov_{\nu}T^{(2)\mu\nu}\neq 0$.
Both of them are useless in forming conserved current through contractions with
Killing or conformal Killing vector fields.
%
%
%
%%%%%%%%%%%%%%%%%%%%%%%%%%%%%%%%%%%%%%%%%%%%%%%%%%%%%
\subsubsection*{Ex: The Robertson-Walker spacetime} %
%%%%%%%%%%%%%%%%%%%%%%%%%%%%%%%%%%%%%%%%%%%%%%%%%%%%%
%
%
%
We do not write down the currents but only mention that
a metric which have all the ingredients to construct a current like
(\ref{cons-curr-f-symb-2-br})
is the Robertson-Walker metric which in spherical coordinates have the expression
\be
ds^{2}=dt^{2}-a^{2}(t)
\left[
\frac{1}{1-kr^{2}}dr^{2}+r^{2}\left(d\theta^{2}+\sin^{2}\theta d\phi^{2}\right)
\right]
\ee
and when
$k\neq 0$ and $a(t)=ct$
admits three rank two f-symbols
[\refcite{Catalin-f-symb}]
\bea
f^{(1)}_{tt}&=&1\\
f^{(2)}_{rt}&=&\frac{1}{\sqrt{1-kr^{2}}}\cos\theta\qquad
f^{(2)}_{\theta t}=-r\sqrt{1-kr^{2}}\sin\theta\\
f^{(3)}_{\phi t}&=&r^{2}\sin^{2}\theta
\eea
and seven Killing vector fields
\bea
\vec{\xi}^{(1)} &=&
\cos\theta\frac{1}{\sqrt{1-kr^{2}}}\;dr-
r\sin\theta\sqrt{1-kr^{2}}\;d\theta\\
\vec{\xi}^{(2)} &=&
\sin\theta\cos\phi\frac{1}{\sqrt{1-kr^{2}}}\;dr+
r\cos\theta\cos\phi\sqrt{1-kr^{2}}\;d\theta-\nonumber\\
&&r\sin\theta\sin\phi\sqrt{1-kr^{2}}\;d\phi\\
\vec{\xi}^{(3)} &=&
\sin\theta\sin\phi\frac{1}{\sqrt{1-kr^{2}}}\;dr+
r\cos\theta\sin\phi\sqrt{1-kr^{2}}\;d\theta+\nonumber\\
&&r\sin\theta\cos\phi\sqrt{1-kr^{2}}\;d\phi\\
\vec{\xi}^{(4)} &=&r^{2}\cos\phi\;d\theta-
r^{2}\sin\theta\cos\theta\sin\phi\;d\phi\\
\vec{\xi}^{(5)} &=&r^{2}\sin\phi\;d\theta+
r^{2}\sin\theta\cos\theta\cos\phi\;d\phi\\
\vec{\xi}^{(6)} &=&r^{2}\sin^{2}\theta;d\phi\\
\vec{\xi}^{(7)} &=&dt
\eea
The Robertson-Walker metric is conformally flat so the Bel-Robinson tensor
$\tbelrob{}_{\mu\nu\lambda\rho}$
being identically zero cannot be used to construct conserved currents.
However this metric have the special feature that the Bel tensor
$\tbel{}_{\mu\nu\lambda\rho}$
is divergenceless in every index, this property making it a potential candidate for
the construction of the conserved currents by contracting it with the rank two f-symbols
and then with the killing vector fields.
Since we need a totally symmetric and divergenceless tensor and since an
operation of total symmetrization does not destroy the divergenceless properties
of the Bel tensor, what we finally use to construct the conserved currents is
the totally symmetrized Bel tensor
$\tbel{}^{(\mu\nu\lambda\rho)}$
and the conserved currents can be constructed as
\be\label{currents-RW}
j^{(ij)\mu}=\tbel{}^{(\mu\nu\lambda\rho)}f^{(i)}_{\lambda\rho}\xi^{(j)}_{\nu}
\ee
%
%
%
%%%%%%%%%%%%%%%%%%%%%%%%%%%%%%%%%%%%%%%%%%%%%%
\subsubsection*{Ex: The Minkowski spacetime} %
%%%%%%%%%%%%%%%%%%%%%%%%%%%%%%%%%%%%%%%%%%%%%%
%
%
%
In Minkowski spacetime
the superenergy tensor associated with a massive or massless scalar field $\Phi$ given by
[\refcite{Senovilla-2000}]
\be\nonumber
S_{\mu\nu\lambda\rho}=
\devcov_{\mu}\devcov_{\lambda}\Phi\devcov_{\rho}\devcov_{\nu}\Phi+
\devcov_{\mu}\devcov_{\rho}\Phi\devcov_{\nu}\devcov_{\lambda}\Phi-
\ee
\be
-g_{\mu\nu}\devcov_{\lambda}\devcov^{\alpha}\Phi\devcov_{\rho}\devcov_{\alpha}\Phi-
g_{\lambda\rho}\devcov_{\mu}\devcov^{\alpha}\Phi\devcov_{\nu}\devcov_{\alpha}\Phi+
\frac{1}{2}g_{\mu\nu}g_{\lambda\rho}
\devcov_{\alpha}\devcov_{\beta}\Phi\devcov^{\alpha}\devcov^{\beta}\Phi
\ee
have the following symmetries
\be
S_{\mu\nu\lambda\rho}=S_{(\mu\nu)(\lambda\rho)}=S_{\lambda\rho\mu\nu}
\ee
and is divergenceless
\be
\devcov_{\alpha}S^{\alpha}_{~\nu\lambda\rho}=0.
\ee
It is easy to see that Minkowski spacetime admits the rank two f-symbols
\be\label{f-symb-r-2-mink}
\fsymbtwo_{\mu\nu}=A_{\mu\nu\alpha}x^{\alpha}+B_{\mu\nu}
\ee
where $A_{\mu\nu\lambda}$ and $B_{\mu\nu}$ are arbitrary constant matrices
except the antisymmetry condition $A_{\mu\nu\lambda}=A_{\mu[\nu\lambda]}$
and the rank three f-symbols
\be
\fsymbthree_{\mu\nu\lambda}=A_{\mu\nu\lambda\alpha}x^{\alpha}+B_{\mu\nu\lambda}
\ee
where $A_{\mu\nu\lambda\rho}$ and $B_{\mu\nu\lambda}$ are arbitrary
constant matrices except the condition $A_{\mu\nu\lambda\rho}=A_{\mu\nu[\lambda\rho]}$,
and also admit the maximum number of ten Killing vector fields.
We can thus construct conserved currents using either rank three f-symbols as in
(\ref{j-fsymb3-bel-def})
\be
j^{\mu}(S,\:h)=
S^{\mu\nu\lambda\rho}[A_{\nu\lambda\rho\alpha}x^{\alpha} + B_{\nu\lambda\rho}]
\ee
either rank two f-symbols as in (\ref{cons-curr-f-symb-2-br})
(with the difference that we use $S^{(\mu\nu\lambda\rho)}$
instead of $S^{\mu\nu\lambda\rho}$ since this operation preserves
divergence free property)
\be
j^{\mu}(S,\:\fsymbtwo,\:\xi)=
S^{(\mu\nu\lambda\rho)}[A_{\lambda\rho\alpha}x^{\alpha}+B_{\lambda\rho}]\xi_{\nu}
\ee
where $\xi_{\nu}$ is one of the ten Killing vector fields in Minkowski spacetime.
%
%
%
%%%%%%%%%%%%%%%%%%%%%%%%%%%%%%%%%%%%%%%%%%%%%%%%%%%%%
\subsection{Killing tensors and conserved currents} %
%%%%%%%%%%%%%%%%%%%%%%%%%%%%%%%%%%%%%%%%%%%%%%%%%%%%%
%
%
%
The generalization of Killing vectors to symmetric tensor fields lead to
the concept of St\"{a}ckel-Killing tensor also known simply as Killing tensors.
A rank $r$ tensor
$K_{\mu_{1}\dots\mu_{r}}$
is a St\"{a}ckel-Killing tensor if it is totally symmetric
$K_{(\mu_{1}\dots\mu_{r})}=K_{\mu_{1}\dots\mu_{r}}$
and satisfy
$$
\devcov_{(\nu}K_{\mu_{1}\dots\mu_{r})}=0.
$$
Rank two St\"{a}ckel-Killing tensors lead to geodesic constants of motion quadratic in momenta
[\refcite{Walker-Penrose-CMP-18-1970}]
and generally higher rank St\"{a}ckel-Killing tensors lead to a corresponding
degree polynomial constants of motion.
The existence of St\"{a}ckel-Killing tensor fields lead to the integrability of geodesic motion
and also to the separability of the Hamilton-Jacobi
[\refcite{Beneti-RMP-12-1977}]
and the Klein-Gordon
[\refcite{Carter-PRD-16-1977}]
equations.

There is a dual relation between the geometry of a space admitting a rank two
St\"{a}ckel-Killing tensor field and the geometry of a space whose metric is that
Killing tensor
[\refcite{Rietdijk-vanHolten-NP-472-1996},
\refcite{Baleanu-Karasu-MPLA-14-1999},
\refcite{Baleanu-Baskal-MPLA-16-2001}
\refcite{Baleanu-IJMP-11-2002}].

St\"{a}ckel-Killing tensors can be also used in conjunction with
Bel and Bel-Robinson type tensors in order to obtain conserved currents.
Thus, instead of a rank three f-symbol a rank three Killing tensor
$\killingthree{}_{\mu\nu\lambda}$
can be used in (\ref{j-3-def}) to obtain the current
\be\label{j-3-def-with-killing}
\jbelrobk{\mu}=\tbelrobtype{}^{\mu\nu\lambda\rho}\:\killingthree{}_{\nu\lambda\rho}
\ee
whose divergence is
\be\label{div-br-k-3-1}
\devcov_{\mu}\jbelrobk{\mu}=
\tbelrobtype{}^{\mu\nu\lambda\rho}\:\devcov_{\mu}\killingthree{}_{\nu\lambda\rho}.
\ee
Since the Killing tensors equation
$
\devcov_{(\mu}\killingthree{}_{\nu\lambda\rho)}=0
$
implies
$
\devcov_{\{\mu}\killingthree{}_{\nu\lambda\rho\}}=0,
$
where the curly brackets stands for summation over the cyclical permutations of the enclosed indices,
we have
\be\label{div-br-k-3-2}
\devcov_{\mu}\jbelrobk{\mu}=
\tbelrobtype{}^{\mu\nu\lambda\rho}
(
-\devcov_{\rho}\killingthree{}_{\mu\nu\lambda}-
\devcov_{\lambda}\killingthree{}_{\rho\mu\nu}-
\devcov_{\nu}\killingthree{}_{\lambda\rho\mu}
)
=
-3\tbelrobtype{}^{\mu\nu\lambda\rho}\devcov_{\mu}\killingthree{}_{\nu\lambda\rho}.
\ee
For the last equality we use the fact that
$\tbelrobtype{}_{\mu\nu\lambda\rho}$
being totally symmetric all its contractions with an arbitrary tensor are the same.
From eq. (\ref{div-br-k-3-1}) and (\ref{div-br-k-3-2}) we evidently have
$
\devcov_{\mu}\jbelrobf{\mu}=0.
$

Again, as for the above presented rank three case, instead of o rank two f-symbol a rank two Killing tensor
can be used to produce a conserved current via a subsequent contraction with a Killing or conformal Killing
vector field.
Thus the contraction between a Bel-Robinson type tensor
$
\tbelrobtype{}^{\mu\nu\alpha\beta}
$
and a rank two Killing tensor
$
P_{\mu\nu}
$
lead to
\be
T^{\mu\nu}=\tbelrobtype{}^{\mu\nu\alpha\beta}\:\killingtwo{}_{\alpha\beta}
\ee
which is evidently symmetric and trace free.
It's divergence is
\be\label{div-j-t-k2}
\devcov_{\mu}T^{\mu\nu}=\tbelrobtype{}^{\mu\nu\alpha\beta}\:\devcov_{\mu}\killingtwo{}_{\alpha\beta}
\ee
The defining properties
$
\killingtwo_{\mu\nu}=\killingtwo_{(\mu\nu)}
$
and
$
\devcov_{(\lambda}\killingtwo_{\mu\nu)}=0
$
of the Killing tensors leads to
$
\devcov_{\{\lambda}\killingtwo_{\mu\nu\}}=0
$
which allows us to write for the above equation
\be
\devcov_{\mu}T^{\mu\nu}=-
\tbelrobtype{}^{\mu\nu\alpha\beta}\:\devcov_{\beta}\killingtwo{}_{\mu\alpha}-
\tbelrobtype{}^{\mu\nu\alpha\beta}\:\devcov_{\alpha}\killingtwo{}_{\beta\mu}=
-2\tbelrobtype{}^{\mu\nu\alpha\beta}\:\devcov_{\mu}\killingtwo{}_{\alpha\beta}
\ee
which compared with
(\ref{div-j-t-k2})
means that
$\devcov_{\mu}T^{\mu\nu}=0$.
Having a symmetric, tracefree and divergenceless rank two tensor field,
a conserved current can be constructed as in the case of f-symbols by
contractions with Killing or even conformal Killing vector fields.

Also, because of the tracefree property of the Bel-Robinson tensor in four dimensions
a rank two conformal Killing tensor, i.e. a rank two symmetric tensor
$\confkillingtwo_{\alpha\beta}$
satisfying
\be
\devcov_{\mu}\confkillingtwo_{\alpha\beta}+
\devcov_{\alpha}\confkillingtwo_{\beta\mu}+
\devcov_{\beta}\confkillingtwo_{\mu\alpha}
=
\chi_{\alpha}g_{\beta\mu}+
\chi_{\beta}g_{\mu\alpha}+
\chi_{\mu}g_{\alpha\beta}
\ee
can be use instead of a St\"{a}ckel-Killing one.
We have for the divergence of the symmetric tensor $T^{\mu\nu}=\tbelrobtype{}^{\mu\nu\alpha\beta}\:\confkillingtwo{}_{\alpha\beta}$
\be\label{eqq}
\devcov_{\mu}T^{\mu\nu}=\tbelrobtype{}^{\mu\nu\alpha\beta}\:\devcov_{\mu}\confkillingtwo{}_{\alpha\beta}
\ee
Since
$
\devcov_{\mu}\confkillingtwo_{\alpha\beta}=
-\devcov_{\alpha}\confkillingtwo_{\beta\mu}-\devcov_{\beta}\confkillingtwo_{\mu\alpha}+
\chi_{\{\alpha}g_{\beta\mu\}}
$
and because of the total symmetry of
$\tbelrobtype{}^{\mu\nu\lambda\rho}$
we have
\be
\devcov_{\mu}T^{\mu\nu}=
-2\tbelrobtype{}^{\mu\nu\alpha\beta}\:\devcov_{\mu}\confkillingtwo{}_{\alpha\beta}+
\tbelrobtype{}^{\mu\nu\alpha\beta}\chi_{\{\alpha}g_{\beta\mu\}}
\ee
As the last term $B^{\mu\nu\alpha\beta}\chi_{\{\alpha}g_{\beta\mu\}}$ is null
because implies various traces of $\tbelrobtype{}^{\mu\nu\alpha\beta}$ which are null by definition, what remains is
\be\label{eq2q}
\devcov_{\mu}T^{\mu\nu}=-2\tbelrobtype{}^{\mu\nu\alpha\beta}\:\devcov_{\mu}\confkillingtwo{}_{\alpha\beta}
\ee
From eq. (\ref{eqq}) and (\ref{eq2q}) we evidently have
$\devcov_{\mu}T^{\mu\nu}=0$ q.e.d..
%
%
%
%%%%%%%%%%%%%%%%%%%%%%%%%%%%%%%%%%%%%%%%%%%%%%%%%%%%%%%%%%
\subsection{Killing-Yano tensors and conserved currents} %
%%%%%%%%%%%%%%%%%%%%%%%%%%%%%%%%%%%%%%%%%%%%%%%%%%%%%%%%%%
%
%
%
The last generalization of Killing vectors we consider refers to Killing-Yano tensors
[\refcite{Yano-AM-55-1952}].
A rank two Killing-Yano tensor is an antisymmetric tensor
$Y_{\mu\nu}=Y_{\nu\mu}$
which satisfies
\be
\devcov_{\lambda}Y_{\mu\nu}+\devcov_{\nu}Y_{\mu\lambda}=0.
\ee
Killing-Yano tensors have major impact in physics.
They play an important role in the existence of the geodesic constants of motion
[\refcite{Carter-PRD-16-1977}]
in curved spacetimes.
There is a deep connection between Killing-Yano tensors,
supersymmetries in pseudo-classical spinning particle models
[\refcite{Gibbons-Rietdijk-vanHolten-NPB-404-1993},
\refcite{vanHolten-NPPS-49-1996},
\refcite{Visinescu-NPPS-56B-1997}]
(models which besides usual spacetimes coordinates includes a number of anticommuting ones to
describe the spin degrees of freedom) and Dirac-type operators on curved spacetimes
[\refcite{Klishevich-CQG-17-2000},
\refcite{Cariglia-CQG-21-2004},
\refcite{Cotaescu-Visinescu-CQG-21-2004},
\refcite{Cotaescu-Visinescu-FP-54-2006}].
Killing-Yano tensors are directly related with the absence of gravitational quantum anomalies
[\refcite{Carter-McLenaghan-PRD-19-1979},
\refcite{Cariglia-CQG-21-2004},
\refcite{Cotaescu-Moroianu-Visinescu-JPA-2005}]
(conservation laws valid at classical level ceasing to be true at quantum level).
Interesting applications of Killing-Yano tensors
like their relations with Nambu tensors and superintegrability can be found in
[\refcite{Baleanu-NCB-114-1999},
\refcite{Defterli-Baleanu-CJP-54-2004}].

There are two conserved currents that can be obtained from a Killing-Yano tensor
$Y_{\mu\nu}$
and a divergenceless Bel-Robinson type tensor
$\tbelrobtype_{\mu\nu\lambda\rho}$.
The first one have the expression
\be\label{cons-curr-ky-001}
\jbelrobky{(1)\mu}=\tbelrobtype{}^{\mu}_{~\alpha_{1}\alpha_{2}\alpha_{3}}
\prod_{i=1}^{3}\varepsilon^{\alpha_{i}\beta\gamma\delta}\devcov_{\delta}Y_{\beta\gamma}
\ee
while for the second we have
\be\label{cons-curr-ky-002}
\jbelrobky{(2)\mu}=\tbelrobtype{}^{\mu}_{~\alpha_{1}\alpha_{2}\alpha_{3}}
\prod_{i=1}^{3}Y^{\alpha_{i}}_{~\gamma}Y^{\gamma}_{~\beta}
\varepsilon^{\beta\delta\sigma\tau}\devcov_{\tau}Y_{\delta\sigma}.
\ee
Neither
$\jbelrobky{(1)\mu}$
nor
$\jbelrobky{(2)\mu}$
are new conserved currents since if a spacetime admit a Killing-Yano tensor
$Y_{\mu\nu}$
than
\be
v^{\mu}=\varepsilon^{\mu\alpha\beta\gamma}\devcov_{\gamma}Y_{\alpha\beta}
\ee
and
\be
w^{\mu}=Y^{\mu}_{~\alpha}Y^{\alpha}_{~\beta}v^{\beta}
\ee
are both Killing vector fields
[\refcite{Carter-JMP-28-1987}].

It is therefore evident that instead of the product
$\prod_{i=1}^{3}\varepsilon^{\alpha_{i}\beta\gamma\delta}\devcov_{\delta}Y_{\beta\gamma}$
in (\ref{cons-curr-ky-001}) we can use a expression like
$
(\varepsilon^{\alpha_{1}\beta\gamma\delta}\devcov_{\delta}Y_{\beta\gamma})
(\varepsilon^{\alpha_{2}\beta\gamma\delta}\devcov_{\delta}Y_{\beta\gamma})
(\varepsilon^{\alpha_{3}\beta\gamma\delta}\devcov_{\delta}Y_{\beta\gamma})
$
since each parenthesis represents in fact a different Killing vector fields.
A similar statement stands true for eq. (\ref{cons-curr-ky-002}).
Also, in (\ref{cons-curr-ky-001}) and (\ref{cons-curr-ky-002}) a symmetrized Bel type tensor
$\tbeltype_{(\mu\nu\lambda\rho)}$
can be used instead of
$\tbelrobtype_{\mu\nu\lambda\rho}$
because a Bel type tensor have only the partial symmetry
$\tbeltype_{\mu\nu\lambda\rho}=\tbeltype_{(\mu\nu)(\lambda\rho)}$
while a symmetry in at least three indices is required to obtain a conserved current.
%
%
%
%%%%%%%%%%%%%%%%%%%%%%%
\section{Conclusions} %
%%%%%%%%%%%%%%%%%%%%%%%
%
%
%
We use three types of generalizations of Killing vector fields -
a symmetrical one which lead to St\"{a}ckel-Killing tensors, an antisymmetrical one
leading to the Killing-Yano tensors and a generalization that do not required
a definite symmetry leading to f-symbols - to construct conserved Bel type currents
by suitable contractions with tensors having the symmetry and divergence properties
of the Bel and Bel-Robinson tensors in Einstein spacetimes. Such a tensor could be for example the basic
superenergy tensor of the scalar field in the Minkowski spacetime.
When the rank four tensor is also trace free a conformal Killing
tensor can be use instead of a St\"{a}ckel-Killing one.
The conserved currents derived by using Killing-Yano tensors are not new,
they could be obtained directly from some Killing vector fields.
As examples we mention Robertson-Walker and Minkowski spacetimes.
We also found the rank two and three f-symbols for the Minkowski spacetime
and write down an integrability condition for rank two f-symbols.
%
%
%
%%%%%%%%%%%%%%%%%%%%%%%%%%%%
\section*{Acknowledgments} %
%%%%%%%%%%%%%%%%%%%%%%%%%%%%
%
%
%
This work was supported by the CNCSIS Program IDEI 571/2008 and NUCLEU 09.39.06.04.


\begin{thebibliography}{99}
\bibitem{Bonilla-Senovilla-1997}
M.A.G. Bonilla and J.M.M. Senovilla,
%{\it Some properties of the Bel and Bel-Robinson tensors},
{\it Gen. Rel. Grav.} {\bf 29}, 91--116 (1997).
%
\bibitem{Senovilla-2000}
J.M.M. Senovilla,
%{\it Super-energy tensors},
{\it Class. Quant. Grav.} {\bf 17}, 2799--2842 (2000).
%
\bibitem{Bel-1959}
L. Bel,
%{\it Introduction d'un tenseur du quatri`eme ordre},
{\it C.R. Acad Sci. Paris} {\bf 248}, 1297--1300 (1959).
%
\bibitem{Bel-1958}
L. Bel,
%{\it Sur la radiation gravitationnelle},
{\it C.R. Acad Sci. Paris} {\bf 247}, 1094--1096 (1958).
%
\bibitem{Saha-Rikhvitsky-Visinescu-MPLA-21-2006}
B. Saha, V. Rikhvitsky and M. Visinescu,
%{\it Bel-Robinson tensor and dominant energy property in the Bianchi type I Universe},
{\it Mod. Phys. Lett. A} {\bf 21}, 847--862 (2006).
%
\bibitem{Lazkoz-Senovilla-Vera-ERE}
R. Lazkoz, J.M.M. Senovilla and R. Vera,
%{\it Properties of Bel currents},
arXiv:gr-qc/0104091.
%
\bibitem{Senovilla-MPLA-2000}
J.M.M. Senovilla,
%{\it (Super)n-Energy for Arbitrary Fields and its Interchange: Conserved Quantities},
{\it Mod. Phys. Lett.} {\bf A15}, 159--166 (2000).
%
\bibitem{Lazkoz-Senovilla-Vera-CQG-2003}
R. Lazkoz, J.M.M. Senovilla and R. Vera,
%{\it Conserved superenergy currents},
{\it Class. Quant. Grav.} {\bf 20}, 4135--4152 (2003).
%
\bibitem{Eriksson-CQG-2006}
 I. Eriksson,
%{\it Conserved Matter Superenergy Currents for Hypersurface Orthogonal Killing Vectors},
{\it Class. Quant. Grav.} {\bf 23}, 2279--2290 (2006).
%
\bibitem{Eriksson-CQG-2007}
I. Eriksson,
%{\it Conserved Matter Superenergy Currents for Orthogonally Transitive Abelian $G_2$ Isometry Groups},
{\it Class. Quant. Grav.} {\bf 24}, 4955--4968 (2007).
%
\bibitem{Gibbons-Rietdijk-vanHolten-NPB-404-1993}
G.W. Gibbons, R.H. Rietdijk and J.W. van Holten,
%{\it SUSY in the Sky},
{\it Nucl.Phys. B} {\bf 404}, 42--64 (1993).
%
\bibitem{Cotaescu-Visinescu-f-symb}
%{\it Non-existence of f-symbols in generalized Taub-NUT spacetime},
I.I. Cotaescu and M. Visinescu,
{\it J. Phys. A: Math. Gen.} {\bf 34}, 6459--6464 (2001).
%
\bibitem{Bergqvist-Lankinen-2004}
G. Bergqvist and P. Lankinen,
%{\it Unique characterization of the Bel–Robinson tensor},
{\it Class. Quant. Grav.} {\bf 21}, 3499--3503 (2004).
\bibitem{Catalin-f-symb}
%
%{\it f-symbols in Robertson-Walker space-time},
F.C. Popa and O. Tintareanu-Mircea,
{\it Centr. Eur. J. Phys.} {\bf 3}, 221--228 (2005).
%
\bibitem{Walker-Penrose-CMP-18-1970}
M. Walker and R. Penrose,
%{\it  On quadratic first integrals of the geodesic equations for type $\{22\}$ spacetimes},
{\it Commun. Math. Phys.} {\bf 18}, 265--274 (1970).
%
\bibitem{Beneti-RMP-12-1977}
%{\it Separable dynamical systems: Characterization of separability structures on Riemannian manifolds},
S. Beneti,
{\it Rep. Math. Phys.} {\bf 12}, 311--316 (1977).
%
\bibitem{Carter-PRD-16-1977}
B. Carter,
%{\it Killing tensor quantum numbers and conserved currents in curved space},
{\it Phys. Rev. D} {\bf 16}, 3395--3414 (1977).
%
\bibitem{Tachibana-TMJ-20-1968}
Shun-ichi Tachibana,
%{\it On Killing tensors in a Riemannian space},
Tohoku Math. J. {\bf 20}, 257---264 (1968).
%
\bibitem{Rietdijk-vanHolten-NP-472-1996}
R.H. Rietdijk and J.W. van Holten,
%{\it Killing tensors and a new geometric duality},
{\it Nucl. Phys. B} {\bf 472}, 427--446 (1996).
%
\bibitem{Baleanu-Karasu-MPLA-14-1999}
D. Baleanu and A. Karasu,
%{\it Lax Tensors, Killing Tensors and Geometric Duality},
{\it Mod. Phys. Lett. A} {\bf 14} 2587--2594 (1999).
%
\bibitem{Baleanu-Baskal-MPLA-16-2001}
%{\it Dual Metrics for a Class of Radiative Spacetimes},
D. Baleanu and S. Baskal,
{\it Mod. Phys. Lett. A} {\bf 16}, 135--142 (2001).
%
\bibitem{Baleanu-IJMP-11-2002}
D. Baleanu,
%{\it Symmetries of the dual metrics},
{\it Int. J. Mod. Phys.} {\bf 11}, 405--416 (2002).
%
\bibitem{Yano-AM-55-1952}
K. Yano,
%{\it Some remarks on tensor fields and curvature},
{\it Ann. Math.} {\bf 55}, 328--347 (1952).
%
\bibitem{vanHolten-NPPS-49-1996}
J.W. van Holten,
%{\it Fermions and world-line supersymmetry},
{\it Nucl. Phys. Proc. Suppl.} {\bf 49}, 319--325 (1996).
%
\bibitem{Visinescu-NPPS-56B-1997}
M. Visinescu,
%{\bf Generalized Killing equations for spinning spaces and the role of Killing-Yano tensors},
{\it Nucl. Phys. Proc. Suppl.} {\bf 56B}, 142--147 (1997).
%
\bibitem{Klishevich-CQG-17-2000}
V.V. Klishevich,
%"On the existence of the second Dirac operator in Riemannian space",
{\it Class. Quant. Grav.} {\bf 17}, 305--318 (2000).
%
\bibitem{Cariglia-CQG-21-2004}
M. Cariglia,
%{\it Quantum mechanics of Yano tensors: Dirac equation in curved spacetimes},
{\it Class. Quant. Grav.} {\bf 21}, 1051--1077 (2004).
%
\bibitem{Cotaescu-Visinescu-CQG-21-2004}
I.I. Cotaescu and M. Visinescu,
%{\it Symmetries of the Dirac operators associated with covariantly constant Killing–Yano tensors},
{\it Class. Quant. Grav.} {\bf 21}, 11--28 (2004).
%
\bibitem{Cotaescu-Visinescu-FP-54-2006}
I.I. Cotaescu and M. Visinescu,
%{\it Superalgebras of Dirac operators on manifolds with special Killing-Yano tensors},
{\it Fortsch. Phys.} {\bf 54}, 1142--1164 (2006).
%
\bibitem{Carter-McLenaghan-PRD-19-1979}
B. Carter and R.G. McLenaghan,
%{\it Generalized total angular momentum operator for the Dirac equation in curved space-time},
{\it Phys. Rev. D} {\bf 19}, 1093--1097 (1979).
%
\bibitem{Cotaescu-Moroianu-Visinescu-JPA-2005}
I.I. Cotaescu, S. Moroianu and M. Visinescu,
%{\it Quantum anomalies for generalized Euclidean Taub-NUT metrics},
{\it J. Phys. A: Math. Gen.} {\bf 38}, 7005--7019 (2005).
%
\bibitem{Baleanu-NCB-114-1999}
D. Baleanu,
%{\it Killing-Yano tensors and Nambu tensors},
{\it Il Nuovo Cimento B} {\bf 114}, 1065--1072 (1999).
%
\bibitem{Defterli-Baleanu-CJP-54-2004}
O. Defterli and D. Baleanu,
%{\it Killing-Yano tensors and superintegrable systems},
{\it Czech. J. Phys.} {\bf 54}, 1215--1221 (2004).
%
\bibitem{Carter-JMP-28-1987}
B. Carter,
%{\it Separability of the Killing–Maxwell system underlying the generalized angular momentum constant in the Kerr–Newman black hole metrics},
{\it J. Math. Phys.} {\bf 28}, 1535--1538 (1987).
%
\end{thebibliography}
\end{document}